# Scalable K-Means++

Bahman Bahmani[*][†]
Stanford University
Stanford, CA
bahman@stanford.edu

Benjamin Moseley[*][‡]
University of Illinois
Urbana, IL
bmosele2@illinois.edu

Andrea Vattani[*][§]
University of California
San Diego, CA
avattani@cs.ucsd.edu

Ravi Kumar
Yahoo! Research
Sunnyvale, CA
ravikumar@yahoo-inc.com

Sergei Vassilvitskii
Yahoo! Research
New York, NY
sergei@yahoo-inc.com

## ABSTRACT

Over half a century old and showing no signs of aging, $k$-means remains one of the most popular data processing algorithms. As is well-known, a proper initialization of $k$-means is crucial for obtaining a good final solution. The recently proposed $k$-means++ initialization algorithm achieves this, obtaining an initial set of centers that is provably close to the optimum solution. A major downside of the $k$-means++ is its inherent sequential nature, which limits its applicability to massive data: one must make $k$ passes over the data to find a good initial set of centers. In this work we show how to drastically reduce the number of passes needed to obtain, in parallel, a good initialization. This is unlike prevailing efforts on parallelizing $k$-means that have mostly focused on the post-initialization phases of $k$-means. We prove that our proposed initialization algorithm $k$-means|| obtains a nearly optimal solution after a logarithmic number of passes, and then show that in practice a constant number of passes suffices. Experimental evaluation on real-world large-scale data demonstrates that $k$-means|| outperforms $k$-means++ in both sequential and parallel settings.

## 1. INTRODUCTION

Clustering is a central problem in data management and has a rich and illustrious history with literally hundreds of different algorithms published on the subject. Even so, a single method — $k$-means — remains the most popular clustering method; in fact, it was identified as one of the top 10 algorithms in data mining [34]. The advantage of $k$-means is its simplicity: starting with a set of randomly chosen initial centers, one repeatedly assigns each input point to its nearest center, and then recomputes the centers given the point assignment. This local search, called *Lloyd's* iteration, continues until the solution does not change between two consecutive rounds.

The $k$-means algorithm has maintained its popularity even as datasets have grown in size. Scaling $k$-means to massive data is relatively easy due to its simple iterative nature. Given a set of cluster centers, each point can independently decide which center is closest to it and, given an assignment of points to clusters, computing the optimum center can be done by simply averaging the points. Indeed parallel implementations of $k$-means are readily available (see, for example, cwiki.apache.org/MAHOUT/k-means-clustering.html).

From a theoretical standpoint, $k$-means is not a good clustering algorithm in terms of efficiency or quality: the running time can be exponential in the worst case [32, 4] and even though the final solution is locally optimal, it can be very far away from the global optimum (even under repeated random initializations). Nevertheless, in practice the speed and simplicity of $k$-means cannot be beat. Therefore, recent work has focused on improving the initialization procedure: deciding on a better way to initialize the clustering dramatically changes the performance of the Lloyd's iteration, both in terms of quality and convergence properties.

A important step in this direction was taken by Ostrovsky et al. [30] and Arthur and Vassilvitskii [5], who showed a simple procedure that both leads to good theoretical guarantees for the quality of the solution, and, by virtue of a good starting point, improves upon the running time of Lloyd's iteration in practice. Dubbed $k$-means++, the algorithm selects only the first center uniformly at random from the data. Each subsequent center is selected with a probability proportional to its contribution to the overall error given the previous selections (we make this statement precise in Section 3). Intuitively, the initialization algorithm exploits the fact that a good clustering is relatively spread out, thus when selecting a new cluster center, preference should be given to those further away from the previously selected centers. Formally, one can show that $k$-means++ initialization leads to an $O(\log k)$ approximation of the optimum [5], or a constant approximation if the data is known to be well-clusterable [30]. The experimental evaluation of $k$-means++ initialization and the variants that followed [1, 2, 15] demonstrated that correctly initializing Lloyd's iteration is crucial if one were to obtain a good solution not only in theory, but also in practice. On a variety of datasets $k$-means++ initialization obtained order of magnitude improvements over the random initialization.

---

[*]Part of this work was done while the author was visiting Yahoo! Research.
[†]Research supported in part by William R. Hewlett Stanford Graduate Fellowship, and NSF awards 0915040 and IIS-0904325.
[‡]Partially supported by NSF grant CCF-1016684.
[§]Partially supported by NSF grant #0905645.

The downside of the $k$-means++ initialization is its inherently sequential nature. Although its total running time of $O(nkd)$, when looking for a $k$-clustering of $n$ points in $\mathbb{R}^d$, is the same as that of a single Lloyd's iteration, it *is not* apparently parallelizable. The probability with which a point is chosen to be the $i$th center depends critically on the realization of the previous $i-1$ centers (it is the previous choices that determine which points are away in the current solution). A naive implementation of $k$-means++ initialization will make $k$ passes over the data in order to produce the initial centers.

This fact is exacerbated in the massive data scenario. First, as datasets grow, so does the number of classes into which one wishes to partition the data. For example, clustering millions of points into $k = 100$ or $k = 1000$ is typical, but a $k$-means++ initialization would be very slow in these cases. This slowdown is even more detrimental when the rest of the algorithm (i.e., Lloyd's iterations) can be implemented in a parallel environment like MapReduce [13]. For many applications it is desirable to have an initialization algorithm with similar guarantees to $k$-means++ and that can be efficiently parallelized.

## 1.1 Our contributions

In this work we obtain a parallel version of the $k$-means++ initialization algorithm and empirically demonstrate its practical effectiveness. The main idea is that instead of sampling a single point in each pass of the $k$-means++ algorithm, we sample $O(k)$ points in each round and repeat the process for approximately $O(\log n)$ rounds. At the end of the algorithm we are left with $O(k \log n)$ points that form a solution that is within a constant factor away from the optimum. We then recluster these $O(k \log n)$ points into $k$ initial centers for the Lloyd's iteration. This initialization algorithm, which we call $k$-means∥, is quite simple and lends itself to easy parallel implementations. However, the analysis of the algorithm turns out to be highly non-trivial, requiring new insights, and is quite different from the analysis of $k$-means++.

We then evaluate the performance of this algorithm on real-world datasets. Our key observations in the experiments are:

- $O(\log n)$ iterations is not necessary and after as little as five rounds, the solution of $k$-means∥ is consistently as good or better than that found by any other method.

- The parallel implementation of $k$-means∥ is much faster than existing parallel algorithms for $k$-means.

- The number of iterations until Lloyd's algorithm converges is smallest when using $k$-means∥ as the seed.

## 2. RELATED WORK

Clustering problems have been frequent and important objects of study for the past many years by data management and data mining researchers.[1] A thorough review of the clustering literature, even restricted to the work in the database area, is far beyond the scope of this paper; the readers are referred to the plethora of surveys available [8, 10, 25, 21, 19]. Below, we only discuss the highlights directly relevant to our work.

Recall that we are concerned with $k$-partition clustering: given a set of $n$ points in Euclidean space and an integer $k$, find a partition of these points into $k$ subsets, each with a representative, also known as a center. There are three common formulations of $k$-partition clustering depending on the particular objective used: $k$-center, where the objective is to minimize the maximum distance between a point and its nearest cluster center, $k$-median, where the objective is to minimize the sum of these distances, and $k$-means, where the objective is to minimize the sum of squares of these distances. All three of these problems are NP-hard, but a constant factor approximation is known for them.

The $k$-means algorithms have been extensively studied from database and data management points of view; we discuss some of them. Ordonez and Omiecinski [29] studied efficient disk-based implementation of $k$-means, taking into account the requirements of a relational DBMS. Ordonez [28] studied SQL implementations of the $k$-means to better integrate it with a relational DBMS. The scalability issues in $k$-means are addressed by Farnstrom et al. [16], who used compression-based techniques of Bradley et al. [9] to obtain a single-pass algorithm. Their emphasis is to initialize $k$-means in the usual manner, but instead improve the performance of the Lloyd's iteration.

The $k$-means algorithm has also been considered in a parallel and other settings; the literature is extensive on this topic. Dhillon and Modha [14] considered $k$-means in the message-passing model, focusing on the speed up and scalability issues in this model. Several papers have studied $k$-means with outliers; see, for example, [22] and the references in [18]. Das et al. [12] showed how to implement EM (a generalization of $k$-means) in MapReduce; see also [36] who used similar tricks to speed up $k$-means. Sculley [31] presented modifications to $k$-means for batch optimizations and to take data sparsity into account. None of these papers focuses on doing a non-trivial initialization. More recently, Ene et al. [15] considered the $k$-median problem in MapReduce and gave a constant-round algorithm that achieves a constant approximation.

The $k$-means algorithms have also been studied from theoretical and algorithmic points of view. Kanungo et al. [23] proposed a local search algorithm for $k$-means with a running time of $O(n^3 \epsilon^{-d})$ and an approximation factor of $9 + \epsilon$. Although the running time is only cubic in the worst case, even in practice the algorithm exhibits slow convergence to the optimal solution. Kumar, Sabharwal, and Sen [26] obtained a $(1 + \epsilon)$-approximation algorithm with a running time linear in $n$ and $d$ but exponential in $k$ and $\frac{1}{\epsilon}$. Ostrovsky et al. [30] presented a simple algorithm for finding an initial set of clusters for Lloyd's iteration and showed that under some data separability assumptions, the algo-rithms achieve an $O(1)$-approximation to the optimum. A similar method, $k$-means++, was independently developed by Arthur and Vassilvitskii [5] who showed that it achieves an $O(\log k)$-approximation but without any assumptions on the data.

---

[1] A paper on database clustering [35] won the 2006 SIGMOD Test of Time Award.

Since then, the $k$-means++ algorithm has been extended to work better on large datasets in the streaming setting. Ailon et al. [2] introduced a streaming algorithm inspired by the $k$-means++ algorithm. Their algorithm makes a single pass over the data while selecting $O(k \log k)$ points and achieves a constant-factor approximation in expectation. Their algorithm builds on an influential paper of Guha et al. [17] who gave a streaming algorithm for the $k$-median problem that is easily adaptable to the $k$-means setting. Ackermann et al. [1] introduced another streaming algorithm based on $k$-means++ and show that it performs well while making a single pass over the input.

We will use MapReduce to demonstrate the effectiveness of our parallel algorithm, but note that the algorithm can be implemented in a variety of parallel computational models. We require only primitive operations that are readily available in any parallel setting. Since the pioneering work by Dean and Ghemawat [13], MapReduce and its open source version, Hadoop [33], has become a de facto standard for large data analysis, and a variety of algorithms have been designed for it [7, 6]. To aid in the formal analysis of MapReduce algorithms, Karloff et al. [24] introduced a model of computation for MapReduce, which has since been used to reason about algorithms for set cover [11], graph problems [27], and other clustering formulations [15].

## 3. THE ALGORITHM

In this section we present our parallel algorithm for initializing Lloyd's iteration. First, we set up some notation that will be used throughout the paper. Next, we present the background on $k$-means clustering and the $k$-means++ initialization algorithm (Section 3.1). Then, we present our parallel initialization algorithm, which we call $k$-means|| (Section 3.3). We present an intuition why $k$-means|| initialization can provide approximation guarantees (Section 3.4); the formal analysis is deferred to Section 6. Finally, we discuss a MapReduce realization of our algorithm (Section 3.5).

### 3.1 Notation and background

Let $X = \{x_1, \ldots, x_n\}$ be a set of points in the $d$-dimensional Euclidean space and let $k$ be a positive integer specifying the number of clusters. Let $||x_i - x_j||$ denote the Euclidean distance between $x_i$ and $x_j$. For a point $x$ and a subset $Y \subseteq X$ of points, the distance is defined as $d(x, Y) = \min_{y \in Y} ||x - y||$. For a subset $Y \subseteq X$ of points, let its *centroid* be given by

$$\mathrm{centroid}(Y) = \frac{1}{|Y|} \sum_{y \in Y} y.$$

Let $\mathcal{C} = \{c_1, \ldots, c_k\}$ be a set of points and let $Y \subseteq X$. We define the *cost* of $Y$ with respect to $\mathcal{C}$ as

$$\phi_Y(\mathcal{C}) = \sum_{y \in Y} d^2(y, \mathcal{C}) = \sum_{y \in Y} \min_{i=1,\ldots,k} ||y - c_i||^2.$$

The goal of $k$-means clustering is to choose a set $\mathcal{C}$ of $k$ centers to minimize $\phi_X(\mathcal{C})$; when it is obvious from the context, we simply denote this as $\phi$. Let $\phi^*$ be the cost of the optimal $k$-means clustering; finding $\phi^*$ is NP-hard [3]. We call a set $\mathcal{C}$ of centers to be an $\alpha$-*approximation* to $k$-means if $\phi_X(\mathcal{C}) \leq \alpha \phi^*$. Note that the centers automatically define a clustering of $X$ as follows: the $i$th cluster is the set of all points in $X$ that are closer to $c_i$ than any other $c_j, j \neq i$.

We now describe the popular method for finding a locally optimum solution to the $k$-means problem. It starts with a random set of $k$ centers. In each iteration, a clustering of $X$ is derived from the current set of centers. The centroids of these derived clusters then become the centers for the next iteration. The iteration is then repeated until a stable set of centers is obtained. The iterative portion of the above method is called *Lloyd's* iteration.

Arthur and Vassilvitskii [5] modified the initialization step in a careful manner and obtained a randomized initialization algorithm called $k$-means++. The main idea in their algorithm is to choose the centers one by one in a controlled fashion, where the current set of chosen centers will stochastically *bias* the choice of the next center (Algorithm 1). The advantage of $k$-means++ is that even the initialization step itself obtains an $(8 \log k)$-approximation to $\phi^*$ in expectation (running Lloyd's iteration on top of this will only improve the solution, but no guarantees can be made). The central drawback of $k$-means++ initialization from a scalability point of view is its inherent *sequential* nature: the choice of the next center depends on the current set of centers.

---

**Algorithm 1** $k$-means++$(k)$ initialization.

1: $\mathcal{C} \leftarrow$ sample a point uniformly at random from $X$
2: **while** $|\mathcal{C}| < k$ **do**
3:     Sample $x \in X$ with probability $\frac{d^2(x,\mathcal{C})}{\phi_X(\mathcal{C})}$
4:     $\mathcal{C} \leftarrow \mathcal{C} \cup \{x\}$
5: **end while**

---

### 3.2 Intuition behind our algorithm

We describe the high-level intuition behind our algorithm. It is easiest to think of random initialization and $k$-means++ initialization as occurring at two ends of a spectrum. The former selects $k$ centers in a single iteration according to a specific distribution, which is the uniform distribution. The latter has $k$ iterations and selects one point in each iteration according to a non-uniform distribution (that is constantly updated after each new center is selected). The provable gains of $k$-means++ over random initialization is precisely in the constantly updated non-uniform selection. Ideally, we would like to achieve the best of both worlds: an algorithm that works in a small number of iterations, selects more than one point in each iteration but in a non-uniform manner, and has provable approximation guarantees. Our algorithm follows this intuition and finds the sweet spot (or the best trade-off point) on the spectrum by carefully defining the number of iterations and the non-uniform distribution itself. While the above idea seems conceptually simple, making it work with provable guarantees (as $k$-means++) throws up a lot of challenges, some of which are also clearly reflected in the analysis of our algorithm. We now describe our algorithm.

## 3.3 Our initialization algorithm: k-means‖

In this section we present k-means‖, our parallel version for initializing the centers. While our algorithm is largely inspired by k-means++, it uses an *oversampling factor* $\ell = \Omega(k)$, which is unlike k-means++; intuitively, $\ell$ should be thought of as $\Theta(k)$. Our algorithm picks an initial center (say, uniformly at random) and computes $\psi$, the initial cost of the clustering after this selection. It then proceeds in $\log \psi$ iterations, where in each iteration, given the current set $\mathcal{C}$ of centers, it samples each $x$ with probability $\ell d^2(x,\mathcal{C})/\phi_X(\mathcal{C})$. The sampled points are then added to $\mathcal{C}$, the quantity $\phi_X(\mathcal{C})$ updated, and the iteration continued. As we will see later,

---
**Algorithm 2** k-means‖ $(k, \ell)$ initialization.
1: $\mathcal{C} \leftarrow$ sample a point uniformly at random from $X$
2: $\psi \leftarrow \phi_X(\mathcal{C})$
3: **for** $O(\log \psi)$ times **do**
4:    $\mathcal{C}' \leftarrow$ sample each point $x \in X$ independently with probability $p_x = \frac{\ell \cdot d^2(x,\mathcal{C})}{\phi_X(\mathcal{C})}$
5:    $\mathcal{C} \leftarrow \mathcal{C} \cup \mathcal{C}'$
6: **end for**
7: For $x \in \mathcal{C}$, set $w_x$ to be the number of points in $X$ closer to $x$ than any other point in $\mathcal{C}$
8: Recluster the weighted points in $\mathcal{C}$ into $k$ clusters

---

the expected number of points chosen in each iteration is $\ell$ and at the end, the expected number of points in $\mathcal{C}$ is $\ell \log \psi$, which is typically more than $k$. To reduce the number of centers, Step 7 assigns weights to the points in $\mathcal{C}$ and Step 8 reclusters these weighted points to obtain $k$ centers. The details are presented in Algorithm 2.

Notice that the size of $\mathcal{C}$ is significantly smaller than the input size; the reclustering can therefore be done quickly. For instance, in MapReduce, since the number of centers is small they can all be assigned to a single machine and any provable approximation algorithm (such as k-means++) can be used to cluster the points to obtain $k$ centers. A MapReduce implementation of Algorithm 2 is discussed in Section 3.5.

While our algorithm is very simple and lends itself to a natural parallel implementation (in $\log \psi$ rounds[2]), the challenging part is to show that it has provable guarantees. Note that $\psi \leq n^2 \Delta^2$, where $\Delta$ is the maximum distance among a pair of points in $X$.

We now state our formal guarantee about this algorithm.

THEOREM 1. *If an $\alpha$-approximation algorithm is used in Step 8, then Algorithm k-means‖ obtains a solution that is an $O(\alpha)$-approximation to k-means.*

Thus, if k-means++ initialization is used in Step 8, then k-means‖ is an $O(\log k)$-approximation. In Section 3.4 we give an intuitive explanation why the algorithm works; we defer the full proof to Section 6.

## 3.4 A glimpse of the analysis

In this section, we present the intuition behind the proof of Theorem 1. Consider a cluster $A$ present in the optimum k-means solution, denote $|A| = T$, and sort the points in $A$ in an increasing order of their distance to centroid$(A)$: let the ordering be $a_1, \ldots, a_T$. Let $q_t$ be the probability that $a_t$ is the first point in the ordering chosen by k-means‖ and $q_{T+1}$ be the probability that no point is sampled from cluster $A$. Letting $p_t$ denote the probability of selecting $a_t$, we have, by definition of the algorithm, $p_t = \ell d^2(a_t, \mathcal{C})/\phi_X(\mathcal{C})$. Also, since k-means‖ picks each point independently, for any $1 \leq t \leq T$, we have $q_t = p_t \prod_{j=1}^{t-1}(1-p_j)$, and $q_{T+1} = 1 - \sum_{t=1}^{T} q_t$.

[2]In practice, our experimental results in Section 5 show that only a few rounds are enough to reach a good solution.

If $a_t$ is the first point in $A$ (w.r.t. the ordering) sampled as a new center, we can either assign all the points in $A$ to $a_t$, or just stick with the current clustering of $A$. Hence, letting

$$s_t = \min \left\{ \phi_A, \sum_{a \in A} ||a - a_t||^2 \right\},$$

we have

$$E[\phi_A(\mathcal{C} \cup \mathcal{C}')] \leq \sum_{t=1}^{T} q_t s_t + q_{T+1} \phi_A(\mathcal{C}),$$

Now, we do a mean-field analysis, in which we assume all $p_t$'s $(1 \leq t \leq T)$ to be equal to some value $p$. Geometrically speaking, this corresponds to the case where all the points in $A$ are very far from the current clustering (and are also rather tightly clustered, so that all $d(a_t, \mathcal{C})$'s $(1 \leq t \leq T)$ are equal). In this case, we have $q_t = p(1-p)^{t-1}$, and hence $\{q_t\}_{1 \leq t \leq T}$ is a monotone decreasing sequence. By the ordering on $a_t$'s, letting

$$s'_t = \sum_{a \in A} ||a - a_t||^2,$$

we have that $\{s'_t\}_{1 \leq t \leq T}$ is an increasing sequence. Therefore

$$\sum_{t=1}^{T} q_t s_t \leq \sum_{t=1}^{T} q_t s'_t \leq \frac{1}{T} \left( \sum_{t=1}^{T} q_t \cdot \sum_{t=1}^{T} s'_t \right),$$

where the last inequality, an instance of Chebyshev's sum inequality [20], is using the inverse monotonicity of sequences $\{q_t\}_{1 \leq t \leq T}$ and $\{s'_t\}_{1 \leq t \leq T}$. It is easy to see that $\frac{1}{T} \sum_{t=1}^{T} s'_t = 2\phi^*_A$. Therefore,

$$E[\phi_A(\mathcal{C} \cup \mathcal{C}')] \leq (1 - q_{T+1}) 2\phi^*_A + q_{T+1} \phi_A(\mathcal{C}).$$

This shows that in each iteration of k-means‖, for each optimal cluster $A$, we remove a fraction of $\phi_A$ and replace it with a constant factor times $\phi^*_A$. Thus, Steps 1–6 of k-means‖ obtain a constant factor approximation to k-means after $O(\log \psi)$ rounds and return $O(\ell \log \psi)$ centers. The algorithm obtains a solution of size $k$ by clustering the chosen centers using a known algorithm. Section 6 contains the formal arguments that work for the general case when $p_t$'s are not necessarily the same.

## 3.5 A parallel implementation

In this section we discuss a parallel implementation of k-means‖ in the MapReduce model of computation. We assume familiarity with the MapReduce model and refer the reader to [13] for further details. As we mentioned earlier, Lloyd's iterations can be easily parallelized in MapReduce and hence, we only focus on Steps 1–7 in Algorithm 2. Step 4 is very simple in MapReduce: each mapper can sample independently and Step 7 is equally simple given a set $\mathcal{C}$ of centers. Given a (small) set $\mathcal{C}$ of centers, computing $\phi_X(\mathcal{C})$ is also easy: each mapper working on an input partition $X' \subseteq X$ can compute $\phi_{X'}(\mathcal{C})$ and the reducer can simply add these values from all mappers to obtain $\phi_X(\mathcal{C})$. This takes care of Step 2 and the update to $\phi_X(\mathcal{C})$ needed for the iteration in Steps 3–6.

Note that we have tacitly assumed that the set $\mathcal{C}$ of centers is small enough to be held in memory or be distributed among all the mappers. While this suffices for nearly all practical settings, it is possible to implement the above steps in MapReduce even without this assumption. Each mapper holding $X' \subseteq X$ and $\mathcal{C}' \subseteq \mathcal{C}$ can output the tuple $\langle x; \arg\min_{c \in \mathcal{C}'} d(x, c) \rangle$, where $x \in X'$ is the key. From this, the reducer can easily compute $d(x, \mathcal{C})$ and hence $\phi_X(\mathcal{C})$. Reducing the amount of intermediate output by the mappers in this case is an interesting research direction.

## 4. EXPERIMENTAL SETUP

In this section we present the experimental setup for evaluating $k$-means∥. The sequential version of algorithms were evaluated on a single workstation with quad-core 2.5GHz processors and 16Gb of memory. The parallel algorithms were run using a Hadoop cluster of 1968 nodes, each with two quad-core 2.5GHz processors and 16GB of memory.

We describe the datasets and baseline algorithms that will be used for comparison.

### 4.1 Datasets

We use three datasets to evaluate the performance of $k$-means∥. The first dataset, GAUSSMIXTURE, is synthetic; a similar version was used in [5]. To generate the dataset, we sampled $k$ centers from a 15-dimensional spherical Gaussian distribution with mean at the origin and variance $R \in \{1, 10, 100\}$. We then added points from Gaussian distributions of unit variance around each center. Given the $k$ centers, this is a mixture of $k$ spherical Gaussians with equal weights. Note that the Gaussians are separated in terms of probability mass — even if only marginally for the case $R = 1$ — and therefore the value of the optimal $k$-clustering can be well approximated using the centers of these Gaussians. The number of sampled points from this mixture of Gaussians is $n = 10,000$.

The other two datasets considered are from real-world settings and are publicly available from the UC Irvine Machine Learning repository (archive.ics.uci.edu/ml/datasets.html). The SPAM dataset consists of 4601 points in 58 dimensions and represents features available to an e-mail spam detection system. The KDDCUP1999 dataset consists of 4.8M points in 42 dimensions and was used for the 1999 KDD Cup. We also used a 10% sample of this dataset to illustrate the effect of different parameter settings.

For GAUSSMIXTURE and SPAM, given the moderate number of points in those datasets, we use $k \in \{20, 50, 100\}$. For KDDCUP1999, we experiment with finer clusterings, i.e., we use $k \in \{500, 1000\}$. The datasets GAUSSMIXTURE and SPAM are studied with the sequential implementation of $k$-means∥, whereas we use the parallel implementation (in the Hadoop framework) for KDDCUP1999.

### 4.2 Baselines

For the rest of the paper, we assume that each initialization method is implicitly followed by Lloyd's iterations. We compare the performance of $k$-means∥ initialization against the following baselines:

- $k$-means++ initialization, as in Algorithm 1;
- Random, which selects $k$ points uniformly at random from the dataset; and
- Partition, which is a recent streaming algorithm for $k$-means clustering [2], described in Section 4.2.1.

Of these, $k$-means++ can be viewed as the true baseline, since $k$-means∥ is a natural parallelization of it. However, $k$-means++ can be only run on datasets of moderate size and only for modest values of $k$. For large-scale datasets, it becomes infeasible and parallelization becomes necessary. Since Random is commonly used and is easily parallelized, we chose it as one of our baselines. Finally, Partition is a recent one-pass streaming algorithm with performance guarantees, and is also parallelizable; hence, we included it as well in our baseline and describe it in Section 4.2.1. We now describe the parameter settings for these algorithms.

For Random, the parallel MapReduce/Hadoop implementation is standard[3]. In the sequential setting we ran Random until convergence, while in the parallel version, we bounded the number of iterations to 20. In general, we observed that the improvement in the cost of the clustering becomes marginal after only a few iterations. Furthermore, taking the best of Random repeated multiple times with different random initial points also obtained only marginal improvements in the clustering cost.

We use $k$-means++ for reclustering in Step 8 of $k$-means∥. We tested $k$-means∥ with $\ell \in \{0.1k, 0.5k, k, 2k, 10k\}$, with $r = 15$ rounds for the case $\ell = 0.1k$, and $r = 5$ rounds otherwise (running $k$-means∥ for five rounds when $\ell = 0.1k$ leads it to select fewer than $k$ centers with high probability). Since the expected number of intermediate points considered by $k$-means∥ is $r\ell$, these settings of the parameters yield a very small intermediate set (of size between $1.5k$ and $40k$). Nonetheless, the quality of the solutions returned by $k$-means∥ is comparable and often better than Partition, which makes use of a much larger set and hence is much slower.

#### 4.2.1 Setting parameters for Partition

The Partition algorithm [2] takes as input a parameter $m$ and works as follows: it divides the input into $m$ equal-sized groups. In each group, it runs a variant of $k$-means++ that selects $3 \log k$ points in each iteration (traditional $k$-means++ selects only a single point). At the end of this, similar to our reclustering step, it runs (vanilla) $k$-means++ on the weighted set of these $3m \log k$ clusters to reduce the number of centers to $k$.

Choosing $m = \sqrt{n/k}$ minimizes the amount of memory used by the streaming algorithm. A neat feature of Partition is that it can be implemented in parallel: in the first round, groups are assigned to $m$ different machines that can be run in parallel to obtain the intermediate set and in the second round, $k$-means++ is run on this set sequentially. In the parallel implementation, the setting $m = \sqrt{n/k}$ not only optimizes the memory used by each machine but also optimizes the total running time of the algorithm (ignoring setup costs), as the size of the instance per machine in the two rounds is equated. (The instance size per machine is $O(n/m) = \tilde{O}(\sqrt{nk})$ which yields a running time in each round of $\tilde{O}(k^{3/2}\sqrt{n})$.) Note that this implies that the running time of Partition does not improve when the number of available machines surpasses a certain threshold. On the other hand, $k$-means‖'s running time improves linearly with the number of available machines (as discussed in Section 2, such issues were considered in [14]). Finally, notice that using this optimal setting, the expected size of the intermediate set used by Partition is $3\sqrt{nk}\log k$, which is much larger than that obtained by $k$-means‖. For instance, Table 5 shows that the size of the coreset returned by $k$-means‖ is smaller by three orders of magnitude.

---

[3]E.g., cwiki.apache.org/MAHOUT/k-means-clustering.html.

|  | $R=1$ | | $R=10$ | | $R=100$ | |
|---|---|---|---|---|---|---|
|  | seed | final | seed | final | seed | final |
| Random | — | 14 | — | 201 | — | 23,337 |
| $k$-means++ | 23 | 14 | 62 | 31 | 30 | 15 |
| $k$-means‖ $\ell=k/2, r=5$ | 21 | 14 | 36 | 28 | 23 | 15 |
| $k$-means‖ $\ell=2k, r=5$ | 17 | 14 | 27 | 25 | 16 | 15 |

**Table 1:** The median cost (over 11 runs) on GAUSS-MIXTURE with $k=50$, scaled down by $10^4$. We show both the cost after the initialization step (*seed*) and the final cost after Lloyd's iterations (*final*).

|  | $k=20$ | | $k=50$ | | $k=100$ | |
|---|---|---|---|---|---|---|
|  | seed | final | seed | final | seed | final |
| Random | — | 1,528 | — | 1,488 | — | 1,384 |
| $k$-means++ | 460 | 233 | 110 | 68 | 40 | 24 |
| $k$-means‖ $\ell=k/2, r=5$ | 310 | 241 | 82 | 65 | 29 | 23 |
| $k$-means‖ $\ell=2k, r=5$ | 260 | 234 | 69 | 66 | 24 | 24 |

**Table 2:** The median cost (over 11 runs) on SPAM scaled down by $10^5$. We show both the cost after the initialization step (*seed*) and the final cost after Lloyd's iterations (*final*).

## 5. EXPERIMENTAL RESULTS

In this section we describe the experimental results based on the setup in Section 4. We present experiments on both sequential and parallel implementations of $k$-means‖. Recall that the main merits of $k$-means‖ were stated in Theorem 1: (i) $k$-means‖ obtains as a solution whose clustering cost is on par with $k$-means++ and hence is expected to be much better than Random and (ii) $k$-means‖ runs in a fewer number of rounds when compared to $k$-means++, which translates into a faster running time especially in the parallel implementation. The goal of our experiments will be to demonstrate these improvements on massive, real-world datasets.

### 5.1 Clustering cost

To evaluate the clustering cost of $k$-means‖, we compare it against the baseline approaches. SPAM and GAUSSMIXTURE are small enough to be evaluated on a single machine, and we compare their cost to that of $k$-means‖ for moderate values of $k \in \{20, 50, 100\}$. We note that for $k \geq 50$, the centers selected by Partition before reclustering represent the full dataset (as $3\sqrt{nk}\log k > n$ for these datasets), which means that results of Partition would be identical to those of $k$-means++. Hence, in this case, we only compare $k$-means‖ with $k$-means++ and Random. KDDCUP1999 is sufficiently large that for large values of $k \in \{500, 1000\}$, $k$-means++ is extremely slow when run on a single machine. Hence, in this case, we will only compare the parallel implementation of $k$-means‖ with Partition and Random.

We present the results for GAUSSMIXTURE in Table 1 and for SPAM in Table 2. For each algorithm we list the cost of the solution both at the end of the initialization step, *before* any Lloyd's iteration and the final cost. We present two parameter settings for $k$-means‖; we will explore the effect of the parameters on the performance of the algorithm in Section 5.3. We note that the initialization cost of $k$-means‖ is typically lower than that of $k$-means++. This suggests that the centers produced by $k$-means‖ avoid outliers, i.e., points that "confuse" $k$-means++. This improvement persists, although is not as pronounced if we look at the final cost of the clustering. In Table 3 we present the results for KDDCUP1999. It is clear that both $k$-means‖ and Partition outperform Random by orders of magnitude. The overall cost for $k$-means‖ improves with larger values of $\ell$ and surpasses that of Partition for $\ell > k$.

|  |  | $k=500$ | $k=1000$ |
|---|---|---|---|
| Random |  | $6.8 \times 10^7$ | $6.4 \times 10^7$ |
| Partition |  | 7.3 | 1.9 |
| $k$-means‖, $\ell=$ | $0.1k$ | 5.1 | 1.5 |
|  | $0.5k$ | 19 | 5.2 |
|  | $k$ | 7.7 | 2.0 |
|  | $2k$ | 5.2 | 1.5 |
|  | $10k$ | 5.8 | 1.6 |

**Table 3:** Clustering cost (scaled down by $10^{10}$) for KDDCUP1999 for $r=5$.

### 5.2 Running time

We now show that $k$-means‖ is faster than Random and Partition when implemented to run in parallel. Recall that the running time of $k$-means‖ consists of two components: the time required to generate the initial solution and the running time of Lloyd's iteration to convergence. The former is proportional to both the number of passes through the data and the size of the intermediate solution.

We first turn our attention to the running time of the initialization routine. It is clear that the number $r$ of rounds used by $k$-means‖ is much smaller than that by $k$-means++. We therefore focus on the parallel implementation and compare $k$-means‖ against Partition and Random. In Table 4 we show the total running time of these algorithms. For various settings of $\ell$, $k$-means‖ runs much faster than Random and Partition.

|  |  | $k=500$ | $k=1000$ |
|---|---|---|---|
| Random |  | 300.0 | 489.4 |
| Partition |  | 420.2 | 1,021.7 |
| $k$-means‖, $\ell=$ | $0.1k$ | 230.2 | 222.6 |
|  | $0.5k$ | 69.0 | 46.2 |
|  | $k$ | 75.6 | 89.1 |
|  | $2k$ | 69.8 | 86.7 |
|  | $10k$ | 75.7 | 101.0 |

**Table 4:** Time (in minutes) for KDDCUP1999.

While one can expect $k$-means‖ to be faster than Random, we investigate the reason why $k$-means‖ runs faster than Partition. Recall that both $k$-means‖ and Partition first select a large number of centers and then recluster the centers to find the $k$ initial points. In Table 5 we show the total number of intermediate centers chosen both by $k$-means‖ and Partition before reclustering on KDDCup1999. We observe that $k$-means‖ is more judicious in selecting centers, and typically selects only 10–40% as many centers as Partition, which directly translates into a faster running time, without sacrificing the quality of the solution. Selecting fewer points in the intermediate state directly translates to the observed speedup.

|  |  | $k = 500$ | $k = 1000$ |
|---|---|---|---|
|  | Partition | $9.5 \times 10^5$ | $1.47 \times 10^6$ |
| $k$-means‖, $\ell =$ | $0.1k$ | 602 | 1,240 |
|  | $0.5k$ | 591 | 1,124 |
|  | $k$ | 1,074 | 2,234 |
|  | $2k$ | 2,321 | 3,604 |
|  | $10k$ | 9,116 | 7,588 |

Table 5: Number of centers for KDDCup1999 before the reclustering.

We next show an unexpected benefit of $k$-means‖: initial solution found by $k$-means‖ leads to a faster convergence of the Lloyd's iteration. In Table 6 we show the number of iterations to convergence of Lloyd's iterations for different initializations. We observe that $k$-means‖ typically requires fewer iterations than $k$-means++ to converge to a local optimum, and both converge significantly faster than Random.

|  | $k = 20$ | $k = 50$ | $k = 100$ |
|---|---|---|---|
| Random | 176.4 | 166.8 | 60.4 |
| $k$-means++ | 38.3 | 42.2 | 36.6 |
| $k$-means‖ $\ell = 0.5k, r = 5$ | 36.9 | 30.8 | 30.2 |
| $k$-means‖ $\ell = 2k, r = 5$ | 23.3 | 28.1 | 29.7 |

Table 6: Number of Lloyd's iterations till convergence (averaged over 10 runs) for Spam.

## 5.3 Trading-off quality with running time

By changing the number $r$ of rounds, $k$-means‖ interpolates between a purely random initialization of $k$-means and the biased sequential initialization of $k$-means++. When $r = 0$ all of the points are sampled uniformly at random, simulating the Random initialization, and when $r = k$, the algorithm updates the probability distribution at every step, simulating $k$-means++. In this section we explore this tradeoff. There is an additional technicality that we must be cognizant of: whereas $k$-means++ draws a single center from the joint distribution induced by $D^2$ weighting, $k$-means‖ selects each point independently with probability proportional to $D^2$, selecting $\ell$ points in *expectation*.

We first investigate the effect of $r$ and $\ell$ on clustering quality. In order to reduce the variance in the computations, and to make sure have exactly $\ell \cdot r$ points at the end of the point selection step, we begin by sampling exactly $\ell$ points from the joint distribution in every round. In Figure 5.1 we show the result on a 10% sample of KDDCup1999, with varying values of $k$

When $\ell = k$ and the algorithm selects exactly $k$ points, we can see that the final clustering cost (after completing the Lloyd's iteration) is monotonically decreasing with the number of rounds. Moreover, even a handful of rounds is enough to substantially bring down the final cost. Increasing $\ell$ to $2k$ and $4k$, while keeping the total number of rounds fixed leads to an improved solution, however this benefit becomes less pronounced as the number of rounds increases. Experimentally we find that the sweet spot lies when $r \approx 8$, and oversampling is beneficial for $r \leq 8$.

In the next set of experiments, we explore the choice of $\ell$ and $r$ when the sampling is done with replacement, as in specifications of $k$-means‖. Recall we can guarantee that the number of rounds needs to be at most $O(\log \psi)$ to achieve a constant competitive solution. However, in practice a smaller number of rounds suffices. (Note that we need at least $k/\ell$ rounds, otherwise we run the risk of having fewer than $k$ centers in the initial set.)

In Figure 5.2 and Figure 5.3, we plot the cost of the final solution as a function of the number of rounds used to initialize $k$-means‖ on GaussMixture and Spam respectively. We also plot the final potential achieved by $k$-means++ as point of comparison. Observe that when $r \cdot \ell < k$, the solution is substantially worse than that of $k$-means++. This is not surprising since in expectation $k$-means‖ has selected too few points. However as soon as $r \cdot \ell \geq k$, the algorithm finds as good of an initial set as that found by $k$-means++.

## 6. ANALYSIS

In this section we present the full analysis of our algorithm, which shows that in each round of the algorithm there is a significant drop in cost of the current solution. Specifically, we show that the cost of the solution drops by a constant factor plus $O(\phi^*)$ in each round — this is the key technical step in our analysis. The formal statement is the following.

THEOREM 2. *Let* $\alpha = \exp\left(-(1 - e^{-\ell/(2k)})\right) \approx e^{-\frac{\ell}{2k}}$. *If* $\mathcal{C}$ *is the set of centers at the beginning of an iteration of Algorithm 2 and* $\mathcal{C}'$ *is the random set of centers added in that iteration, then*

$$E[\phi_X(\mathcal{C} \cup \mathcal{C}')] \leq 8\phi^* + \frac{1+\alpha}{2}\phi_X(\mathcal{C}).$$

Before we proceed to prove Theorem 2, we consider its following simple corollary.

COROLLARY 3. *If* $\phi^{(i)}$ *is the cost of the clustering after the ith round of Algorithm 2, then*

$$E[\phi^{(i)}] \leq \left(\frac{1+\alpha}{2}\right)^i \psi + \frac{16}{1-\alpha}\phi^*.$$

PROOF. By an induction on $i$. The base case $i = 0$ is trivial, as $\phi^{(0)} = \psi$. Assume the claim is valid up to $i$. Then, we will prove it for $i + 1$. From Theorem 2, we know that

$$E[\phi^{(i+1)}|\phi^{(i)}] \leq \frac{1+\alpha}{2} \cdot \phi^{(i)} + 8\phi^*.$$

By taking an expectation over $\phi^{(i)}$, we have

$$E[\phi^{(i+1)}] \leq \frac{1+\alpha}{2}E[\phi^{(i)}] + 8\phi^*.$$

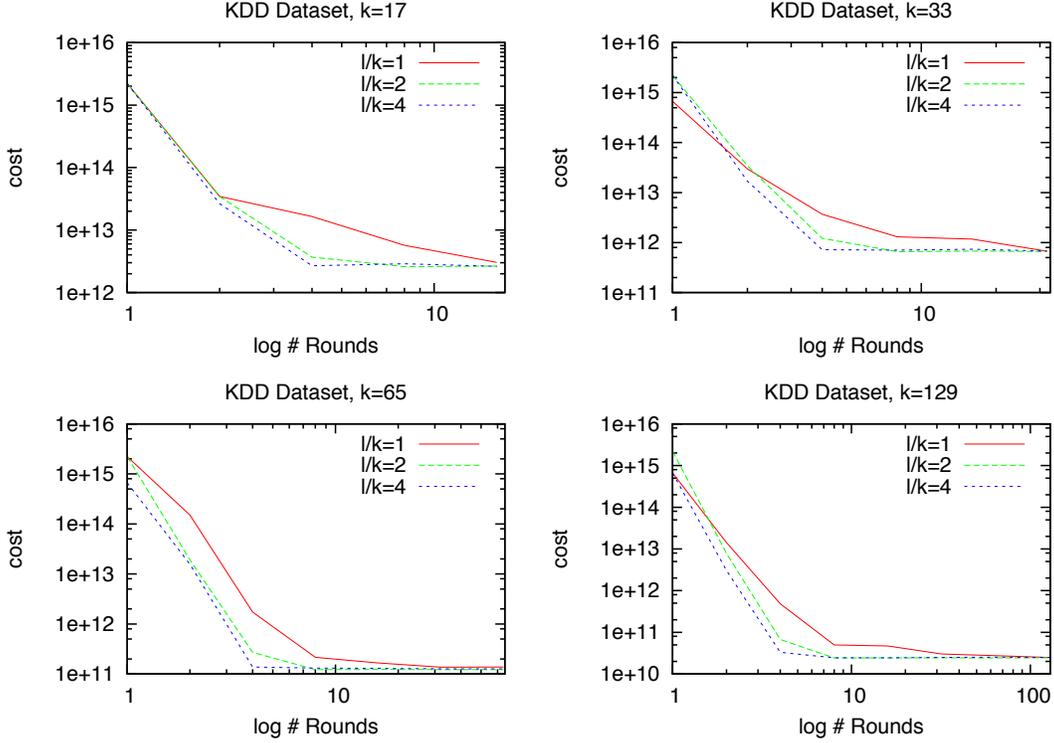

Figure 5.1: The effect of different values of $\ell$ and the number of rounds $r$ on the final cost of the algorithm for a 10% sample of KDDCup1999. Each data point is the median of 11 runs of the algorithm.

By the induction hypothesis on $E[\phi^{(i)}]$, we have

$$E[\phi^{(i+1)}] \leq \left(\frac{1+\alpha}{2}\right)^{i+1}\psi + 8\left(\frac{1+\alpha}{1-\alpha}+1\right)\phi^*$$
$$= \left(\frac{1+\alpha}{2}\right)^{i+1}\psi + \frac{16}{1-\alpha}\phi^*. \quad \square$$

Corollary 3 implies that after $O(\log \psi)$ rounds, the cost of the clustering is $O(\phi^*)$; Theorem 1 is then an immediate consequence. We now proceed to establish Theorem 2.

Consider any cluster $A$ with centroid$(A)$ in the optimal solution. Denote $|A| = T$ and let $a_1, \ldots, a_T$ be the points in $A$ sorted increasingly with respect to their distance to centroid$(A)$. Let $\mathcal{C}'$ denote the set of centers that are selected during a particular iteration. For $1 \leq t \leq T$, we let

$$q_t = \Pr[a_t \in \mathcal{C}', a_j \notin \mathcal{C}', \forall 1 \leq j < t]$$

be the probability that the first $t-1$ points $\{a_1, \ldots, a_{t-1}\}$ are not sampled during this iteration and $a_t$ is sampled. Also, we denote by $q_{T+1}$ the probability that no point is sampled from cluster $A$.

Furthermore, for the remainder of this section, let $D(a) = d(a, \mathcal{C})$, where $\mathcal{C}$ is the set of centers in the current iteration. Letting $p_t$ denote the probability of selecting $a_t$, we have, by definition of the algorithm, $p_t = \frac{\ell D^2(a_t)}{\phi}$. Since $k$-means$\|$ picks each point independently, using the convention that $p_{T+1} = 1$, we have for all $1 \leq t \leq T+1$,

$$q_t = p_t \prod_{j=1}^{t-1}(1 - p_j).$$

The main idea behind the proof is to consider only those clusters in the optimal solution that have significant cost relative to the total clustering cost. For each of these clusters, the idea is to first express both its clustering cost and the probability that an early point is not selected as linear functions of the $q_t$'s (Lemmas 4, 5), and then appeal to linear programming (LP) duality in order to bound the clustering cost itself (Lemma 6 and Corollary 7). To formalize this idea, we start by defining

$$s_t = \min\left\{\phi_A, \sum_{a \in A}||a - a_t||^2\right\},$$

for all $1 \leq t \leq T$, and $s_{T+1} = \phi_A$. Then, letting $\phi'_A = \phi_A(\mathcal{C} \cup \mathcal{C}')$ be the clustering cost of cluster $A$ after the current round of the algorithm, we have the following.

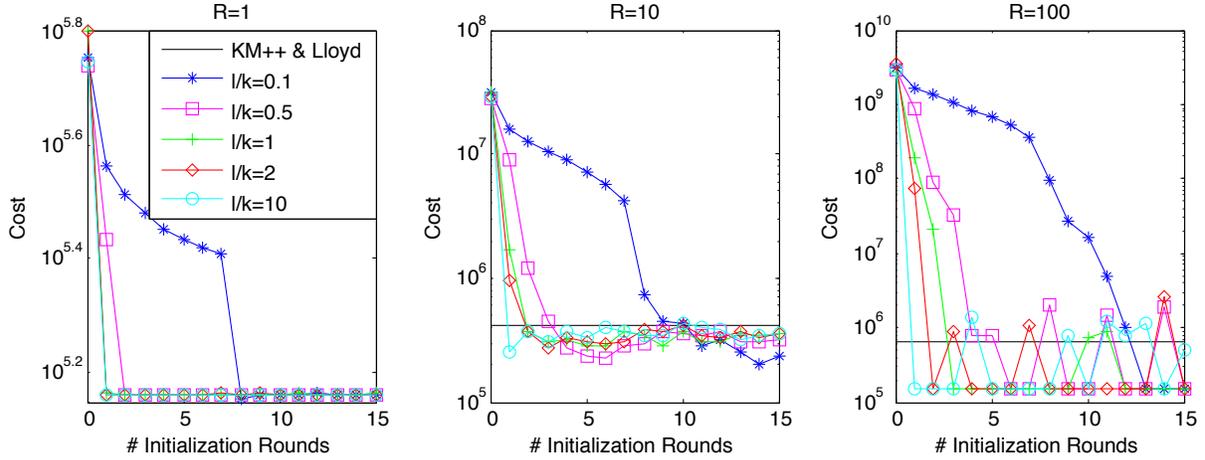

**Figure 5.2:** The cost of $k$-means|| followed by Lloyd's iterations as a function of the number of initialization rounds for GAUSSMIXTURE.

LEMMA 4. *The expected cost of clustering an optimum cluster $A$ after a round of Algorithm 2 is bounded as*

$$E[\phi'_A] \leq \sum_{t=1}^{T+1} q_t s_t. \quad (6.1)$$

PROOF. We can rewrite the expectation of the clustering cost $\phi'_A$ for cluster $A$ after one round of the algorithm as follows:

$$E[\phi'_A] = \sum_{t=1}^{T} q_t E[\phi'_A \mid a_t \in C', a_j \notin C', \forall 1 \leq j < t]. \quad (6.2)$$

Observe that conditioned on the fact that $a_t \in C'$, we can either assign all the points in $A$ to center $a_t$, or just stick with the former clustering of $A$, whichever has a smaller cost. Hence

$$E[\phi'_A \mid a_t \in C', a_j \notin C', \forall 1 \leq j < t] \leq s_t.$$

and the result follows from (6.2). □

In order to minimize the right hand side of (6.1), we want to be sure that the sampling done by the algorithm places a lot of weight on $q_t$ for small values of $t$. Intuitively this means that we are more likely to select a point close to the optimal center of the cluster than one further away. Our sampling based on $D^2(\cdot)$ implies a constraint on the probability that an early point is not selected, which we detail below.

LEMMA 5. *Let $\eta_0 = 1$, and, for any $1 \leq t \leq T$, $\eta_t = \prod_{j=1}^{t}\left(1 - \frac{D^2(a_j)}{\phi_A}(1 - q_{T+1})\right)$. Then, for any $0 \leq t \leq T$,*

$$\sum_{r=t+1}^{T+1} q_r \leq \eta_t.$$

PROOF. First note that $q_{T+1} = \prod_{t=1}^{T}(1 - p_t) \geq 1 - \sum_{t=1}^{T} p_t$. Therefore,

$$1 - q_{T+1} \leq \sum_{t=1}^{t} p_t = \ell \frac{\phi_A}{\phi}.$$

Thus

$$p_t = \frac{\ell D^2(a_t)}{\phi} \geq \frac{D^2(a_t)}{\phi_A}(1 - q_{T+1}).$$

To prove the lemma, by the definition of $q_r$ we have

$$\sum_{r=t+1}^{T+1} q_r = \left(\prod_{j=1}^{t}(1 - p_j)\right) \cdot \sum_{r=t+1}^{T+1} \prod_{j=t+1}^{r-1}(1 - p_j) p_r$$

$$\leq \prod_{j=1}^{t}(1 - p_j)$$

$$\leq \prod_{j=1}^{t}\left(1 - \frac{D^2(a_t)}{\phi_A}(1 - q_{T+1})\right)$$

$$= \eta_t. \quad \square$$

Having proved this lemma, we now slightly change our perspective and think of the values $q_t$ ($1 \leq t \leq T+1$) as variables that (by Lemma 5) satisfy a number of linear constraints and also (by Lemma 4) a linear function of which bounds $E[\phi'_A]$. This naturally leads to an LP on these variables to get an upper bound on $E[\phi'_A]$; see Figure 6.1. We will then use the properties of the LP and its dual to prove the following lemma.

LEMMA 6. *The expected potential of an optimal cluster $A$ after a sampling step in Algorithm 2 is bounded as*

$$E[\phi'_A] \leq (1 - q_{T+1}) \sum_{t=1}^{T} \frac{D^2(a_t)}{\phi_A} s_t + \eta_T \phi_A.$$

PROOF. Since the points in $A$ are sorted increasingly with respect to their distances to the centroid, letting

$$s'_t = \sum_{a \in A} \|a - a_t\|^2, \quad (6.3)$$

for $1 \leq t \leq T$, we have that $s'_1 \leq \cdots \leq s'_T$. Hence, since $s_t = \min\{\phi_A, s'_t\}$, we also have $s_1 \leq \cdots \leq s_T \leq s_{T+1}$.

Now consider the LP in Figure 6.1 and its dual. Since $s_t$ is an increasing sequence, the optimal solution to the dual must have $\alpha_t = s_{t+1} - s_t$ (letting $s_0 = 0$). Then, we can

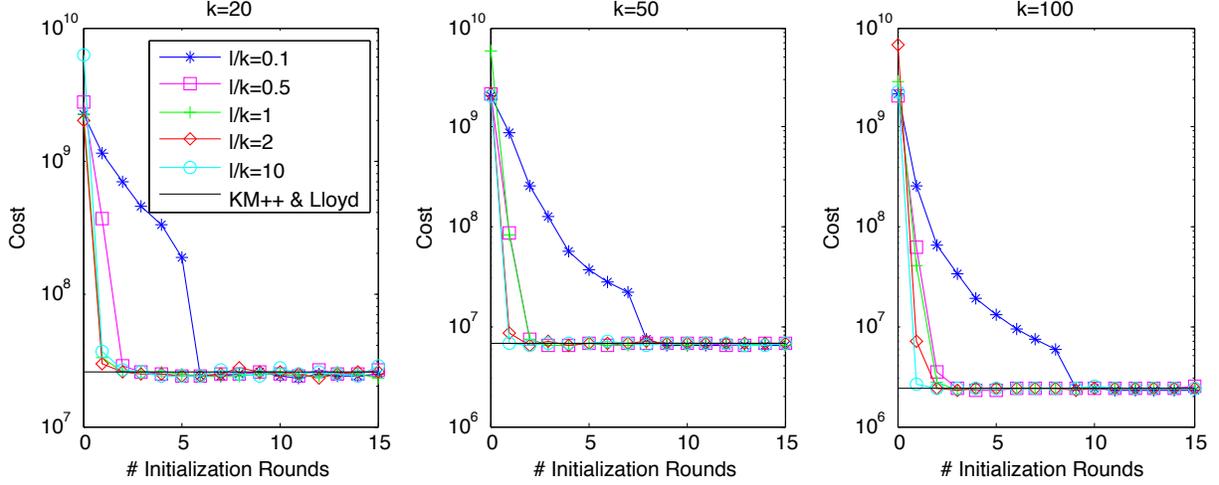

Figure 5.3: The cost of $k$-means|| followed by Lloyd's iterations as a function of the number of initialization rounds for SPAM.

$$\max_{q_1,\ldots,q_{T+1}} \sum_{t=1}^{T+1} q_t s_t$$
subject to
$$\sum_{r=t+1}^{T+1} q_r \leq \eta_t, \quad (\forall 0 \leq t \leq T)$$
$$q_t \geq 0. \quad (\forall 1 \leq t \leq T+1)$$

$$\min_{\alpha_0,\ldots,\alpha_T} \sum_{t=0}^{T} \eta_t \alpha_t$$
subject to
$$\sum_{r=0}^{t-1} \alpha_r \geq s_t, \quad (\forall 1 \leq t \leq T+1)$$
$$\alpha_t \geq 0. \quad (\forall 0 \leq t \leq T)$$

Figure 6.1: LP (left) and its dual (right).

bound the value of the dual (and hence the value of the primal, which by Lemma 4 and Lemma 5 is an upper bound on $E[\phi'_A]$) as follows:

$$E[\phi'_A] \leq \sum_{t=0}^{T} \eta_t \alpha_t$$
$$= \sum_{t=1}^{T} s_t(\eta_{t-1} - \eta_t) + \eta_T s_{T+1}$$
$$= \sum_{t=1}^{T} s_t \eta_{t-1} \left( \frac{D^2(a_t)}{\phi_A} \right)(1 - q_{T+1}) + \eta_T s_{T+1}$$
$$\leq (1 - q_{T+1}) \sum_{t=1}^{T} \frac{D^2(a_t)}{\phi_A} s_t + \eta_T \phi_A,$$

where the last step follows since $\eta_t \leq 1$. □

This results in the following corollary:

COROLLARY 7.
$$E[\phi'_A] \leq 8\phi^*_A(1 - q_{T+1}) + \phi_A e^{-(1-q_{T+1})}.$$

PROOF. By the triangle inequality, for all $a, a_t$ we have $D(a_t) \leq D(a) + ||a - a_t||$. The power-mean inequality then implies that $D^2(a_t) \leq 2D^2(a) + 2||a - a_t||^2$. Summing over all $a \in A$ and dividing by $\phi_A$, we have that

$$\frac{D^2(a_t)}{\phi_A} \leq \frac{2}{T} + \frac{2}{T} \frac{s'_t}{\phi_A},$$

where $s'_t$ is defined in (6.3). Hence,

$$\sum_{t=1}^{T} \frac{D^2(a_t)}{\phi_A} s_t \leq \sum_{t=1}^{T} \left(\frac{2}{T} + \frac{2}{T} \frac{s'_t}{\phi_A}\right) s_t$$
$$= \frac{2}{T} \sum_{t=1}^{T} s_t + \frac{2}{T} \sum_{t=1}^{T} \frac{s'_t s_t}{\phi_A}$$
$$\leq \frac{2}{T} \sum_{t=1}^{T} s'_t + \frac{2}{T} \sum_{t=1}^{T} s'_t$$
$$= 8\phi^*_A,$$

where in the last inequality, we used $s_t \leq s'_t$ for the first summation, and $s_t \leq \phi_A$ for the second one.

Finally, noticing

$$\eta_T = \prod_{j=1}^{T}\left(1 - \frac{D^2(a_j)}{\phi_A}(1 - q_{T+1})\right)$$
$$\leq \exp\left(-\sum_{j=1}^{T}\frac{D^2(a_j)}{\phi_A}(1 - q_{T+1})\right)$$
$$= \exp(-(1 - q_{T+1})),$$

the proof follows from Lemma 6. □

We are now ready to prove the main result.

PROOF OF THEOREM 2. Let $A_1, \ldots, A_k$ be the clusters in the optimal solution $\mathcal{C}_{\phi^*}$. We partition these clusters into "heavy" and "light" as follows:

$$\mathcal{C}_H = \left\{ A \in \mathcal{C}_{\phi^*} \mid \frac{\phi_A}{\phi} > \frac{1}{2k} \right\}, \text{ and}$$

$$\mathcal{C}_L = \mathcal{C}_{\phi^*} \setminus \mathcal{C}_H.$$

Recall that

$$q_{T+1} = \prod_{j=1}^{T}(1 - p_j) \leq \exp\left(-\sum_j p_j\right) = \exp\left(-\frac{\ell\phi_A}{\phi}\right).$$

Then, by Corollary 7, for any heavy cluster $A$, we have

$$E[\phi'_A] \leq 8\phi_A^*(1 - q_{T+1}) + \phi_A e^{-(1-q_{T+1})}$$
$$\leq 8\phi_A^* + \exp(-(1 - e^{-\ell/2k}))\phi_A$$
$$= 8\phi_A^* + \alpha\phi_A.$$

Summing up over all $A \in \mathcal{C}_H$, we get

$$E[\phi'_{\mathcal{C}_H}] \leq 8\phi_{\mathcal{C}_H}^* + \alpha\phi_{\mathcal{C}_H}.$$

Then, by noting that

$$\phi_{\mathcal{C}_L} \leq \frac{\phi}{2k} \cdot |\mathcal{C}_L| \leq \frac{\phi}{2k}k = \frac{\phi}{2},$$

and that $E[\phi'_{\mathcal{C}_L}] \leq \phi_{\mathcal{C}_L}$, we have

$$E[\phi'] \leq 8\phi_{\mathcal{C}_H}^* + \alpha\phi_{\mathcal{C}_H} + \phi_{\mathcal{C}_L}$$
$$= 8\phi_{\mathcal{C}_H}^* + \alpha\phi + (1-\alpha)\phi_{\mathcal{C}_L}$$
$$\leq 8\phi_{\mathcal{C}_H}^* + (\alpha + (1-\alpha)/2)\phi$$
$$\leq 8\phi^* + (\alpha + (1-\alpha)/2)\phi. \quad \Box$$

## 7. CONCLUSIONS

In this paper we obtained an efficient parallel version $k$-means‖ of the inherently sequential $k$-means++. The algorithm is simple and embarrassingly parallel and hence admits easy realization in any parallel computational model. Using a non-trivial analysis, we also show that $k$-means‖ achieves a constant factor approximation to the optimum. Experimental results on large real-world datasets (on which many existing algorithms for $k$-means can grind for a long time) demonstrate the scalability of $k$-means‖.

There have been several modifications to the basic $k$-means algorithm to suit specific applications. It will be interesting to see if such modifications can also be efficiently parallelized.

### Acknowledgments

We thank the anonymous reviewers for their many useful comments.